\documentclass[12pt]{iopart}
\usepackage{graphicx}
\usepackage{xcolor}
\expandafter\let\csname equation*\endcsname=\relax 
\expandafter\let\csname endequation*\endcsname=\relax 
\usepackage{amsmath,amssymb,amsthm} 
\usepackage{braket} 

\begin{document}
\title{The role of stacking on the electronic structure of MoSe$_2$ at small twist angles}

\author{ S. Patra, M. Das and P. Mahadevan}

\address{Departments of Condensed Matter and Material Physics.\\
S. N. Bose National Centre for Basic Sciences, Salt Lake, Kolkata-700106, India}
\ead{priya.mahadevan@gmail.com}
\vspace{10pt}
\begin{indented}
\item[]January 2023
\end{indented}

\begin{abstract}
We consider two high symmetry stackings AA and AB and examine the changes induced in the electronic structure by considering small angles of rotation of 3.48$^{\circ}$ from both these stackings. In both cases we largely recover the low energy electronic structure of the untwisted limit. We additionally find flat bands emerging above the dispersing bands. Surprisingly, while the rotation from the AA end leads to one flat band above the highest occupied band at $\Gamma$, one finds two flat bands emerging for small rotations from the AB end. 
Examining the real space localization of the flat bands allows us to discuss the origin of the flat bands in terms of quantum well states and qualitatively understand the dependence of the number of flat bands found on the twist angle.   
\end{abstract}
%
%
%
%

\section{Introduction}

Advances in experimental techniques have allowed us to assemble van der Waals heterostructures from various layered two dimensional materials. The weak van der Waals interactions that exists between the layers has allowed for an enormous freedom in the construction  of the heterostructures. The layers could be displaced or rotated with respect to each other. The most surprising aspect that has emerged is that these
assembled heterostructures show unique characteristics which are not found in their component materials. This has led to the huge interest in these materials \cite{tw_graphene1,tw_graphene2,tw_graphene3,tw_graphene4,tw-WSe2_1, tw-WSe2_2,tw_graphene5,tw-MoSe2_1}.
 The twisted bilayers were first realised in graphene \cite{tw_graphene1,tw_graphene2,tw_graphene3,tw_graphene4} and were accompanied  by the formation of flat bands and strongly correlated phases. Later on, this has been achieved in layered Mo and W based transition metal dichalcogenides \cite{tw-WSe2_1,tw-WSe2_2,tw-MoSe2_1}. At certain twist angles, these materials exhibit several unusual properties like the formation of flat bands, zero resistive behaviour, electric field modulated metal-insulator transitions etc. which are not found in their untwisted limit \cite{tw-WSe2_1, tw-WSe2_2}. 

Rotating one layer with respect to the other breaks the periodicity of the primitive cell. As a result, a large moir\'{e} cell appears. This leads to a smaller Brillouin zone. Consequently, bands get folded to a significantly small Brillouin zone, emerging as weakly dispersing flat bands. This has been an explanation offered for the formation of flat bands and the associated strong correlation phenomena in these twisted bilayers \cite{tw_theo1,tw_theo2}. Besides, there have been theoretical reports that shows formation of flat bands at certain limits using continuum model \cite{continuum01,continuum02,continuum03}. However, there is no discussion given regarding the mechanism responsible for its formation.

Each layer is coupled to the other by weak van der Waals forces implying weak interactions between the layers. Consequently, twisting one layer with respect to the other is not expected to introduce large changes in the electronic structure compared to the untwisted limit. Therefore, we have labelled the bands of the moir\'{e} cell with the k-points of the untwisted unit cell. In order to do this, we have projected the eigenfunctions of the moir\'{e} bands onto those of the untwisted limit. In case of a perfect supercell this would give us the primitive cell bandstructure. However, for twisted bilayers the unit cells are not perfect supercells. Hence, one would expect that projecting the moir\'{e} cell bandstructure onto the untwisted limit, would provide us with a measure of the extent of perturbation. As the perturbation is expected to be weak, we should largely recover the band structure of the untwisted limit. 

While the unfolding of the electronic structure for large unit cells has been used widely in the context of alloys\cite{alloy+doped, alloy_unfold}, doped unit cells\cite{alloy+doped,doped_unfold}, our earlier work\cite{twist_mose2_pk,twist_mose2_sp,twist_wse2_sp} established that the electronic structure in these van der Waals systems should be viewed as a perturbation to the untwisted limit. In a recent work\cite{twist_mose2_sp}, we have examined the electronic structure of two large and similar sized  moir\'{e} cells for twist angles 19.03$^{\circ}$ and 3.48$^{\circ}$ for MoSe$_2$. By projecting out the eigenfunctions of the bands of the large moir\'{e} cell onto the primitive cell eigenfunctions for a twist angle of 19.03$^{\circ}$, dispersing bands were found similar to the untwisted limit with the electronic structure being hardly perturbed. However, carrying out the same analysis for a twist angle of 3.48$^{\circ}$, we found that in addition to the dispersing bands which can be identified with the untwisted limit, a weakly dispersing band emerges above these bands. This has been associated with the presence of regions of strong perturbation in the moire cell. 

However the effect of rotations upon starting with different stackings has not been adequately examined. We explore this by considering twist angles near 0$^{\circ}$ and 60$^{\circ}$ for AA stacked MoSe$_2$. Here 60$^{\circ}$ rotation from the AA stacking leads to another high symmetry stacking, AB stacking. Carrying out  a similar analysis, we found a pair of split off bands at the top of the valence band. This split-off band  has a vanishingly small bandwidth and is even more flat than what we had for 3.48$^{\circ}$. This has been associated with stronger localisation found here. 
In each of the instances, the
localization of the wavefunction associated with the flat band is in a region in which one 
has an enhanced interlayer hopping interaction. This leads to the antibonding states 
to be pushed to
higher energies 
compared to the states in the region in which the interaction strength is less.
We can associate the 
flat band with the antibonding states arising from these 
interactions. 
The regions characterized by enhanced interlayer interactions can be 
considered to be
associated with a 
potential well of width L.
This mapping allows us to discuss the emergence of the flat bands for a twist angle of 56.52$^{\circ}$  with the solutions of a potential well far separated from its neighbour, allowing us to have two bound states (the flat bands). However, 
for a twist angle of 3.48$^{\circ}$, the solutions are for two potential wells of similar depth and width L, separated by a 
small region. In the 
limit where the separation between them vanishes, we can 
approximate the well width to 2L. As the energy of the solutions varies as $\frac{1}{(2L)^2}$, the shallower energies here allows us to have only one bound state (one flat band).

\section{Methodology}
Ab initio electronic structure calculations have been carried out using density functional theory (DFT) as implemented in the Vienna Ab initio Simulation Package (VASP)\cite{vasp-1,vasp-2}. Projector augmented plane wave potentials (PAW) have been used to describe the ion-electron interactions\cite{paw}. The generalised gradient approximation \cite{pbe-1,pbe-2} has been used for the exchange-correlation functional. The lattice constants of MoSe$_2$ has been kept fixed at the experimental value of  3.289 $\AA$.  Internal positions of the atoms have been optimised through a total energy minimization. A vacuum of 20 $\AA$ has been introduced along the z-direction for all cases to minimise the interaction between the periodic images which are inevitably present because of the periodic implementation of DFT that we use. Van der Waals interactions have been introduced between the two layers using the DFT-D2 method of Grimme\cite{dft-d2}. The electronic structures for twisted bilayers have been solved
self-consistently at $\Gamma$ point. Starting from AA stacking, we rotated the upper layer with respect to the lower one anticlockwise by an angle $\theta$. 
In order to examine the twist induced perturbation, the electronic structures calculated for the large moir\'{e} cell have been projected onto the high-symmetry directions of the unrotated primitive cell using the unfolding technique\cite{unfold}.

A primitive cell wave vector $\textbf{k}_{pr}$ gets folded to a supercell wave vector $\textbf{K}_{sc}$ if there exists a supercell reciprocal lattice translation vector \textbf{G} which satisfies the condition $\textbf{k}_{pr}$ =  $\textbf{K}_{sc}$ +  $\textbf{G}$. Now using Bl\"{o}ch's theorem, any energy eigenstate corresponding to the supercell wave vector $\textbf{K}_{sc}$ could be written as

 \begin{eqnarray}
 \ket{\textbf{K}_{sc}m}=u_{{\textbf{K}_{sc}}m}(r)e^{i\textbf{K}_{sc}\textbf{r}}\label{ksc1}
 \end{eqnarray}
where $u_{{\textbf{K}_{sc}}m}(r)$ is the Bl\"{o}ch function and has the periodicity of the supercell. This can be expanded in terms of the reciprocal lattice vectors $\textbf{G}$ of the supercell as $u_{{\textbf{K}_{sc}}m}(r)$ = $[\sum_{\textbf{G}}\textbf{C}_{\textbf{K}_{sc}m} (\textbf{G})e^{i\textbf{G}\textbf{r}}]$.

So Eqn. \ref{ksc1} can be written as
\begin{eqnarray}
\ket{\textbf{K}_{sc}m}=[\sum_{\textbf{G}}\textbf{C}_{\textbf{K}_{sc}m}(\textbf{G})e^{i\textbf{G}\textbf{r}}]e^{i\textbf{K}_{sc}\textbf{r}}\label{ksc2}
\end{eqnarray}
We determined these Fourier coefficients using the open source interface Wavetrans\cite{wavetrans} for VASP. Then, the contribution of the m-th eigenstate of supercell k-point $\textbf{K}_{sc}$ at the primitive cell k-point $\textbf{k}_{pr}$ is calculated as 
\begin{eqnarray}
P_{\textbf{K}_{sc}m}(k_{pr})=\sum_{n}|\langle\textbf{k}_{pr}n\vert\textbf{K}_{sc}m\rangle|^2=|\textbf{C}_{\textbf{K}_{sc}m}(\textbf{g}+\textbf{k}_{pr}-\textbf{K}_{sc})|^2
\end{eqnarray}
where \textbf{g} is the translation vector of the reciprocal lattice of the primitive cell. Only those $\textbf{G}$ coefficients which are equal to $ \textbf{g}+\textbf{k}_{pr} -\textbf{K}_{sc} $ will contribute. 


\section{Results and Discussions}

\begin{figure}[h!]
\centering
\includegraphics[width=6 in]{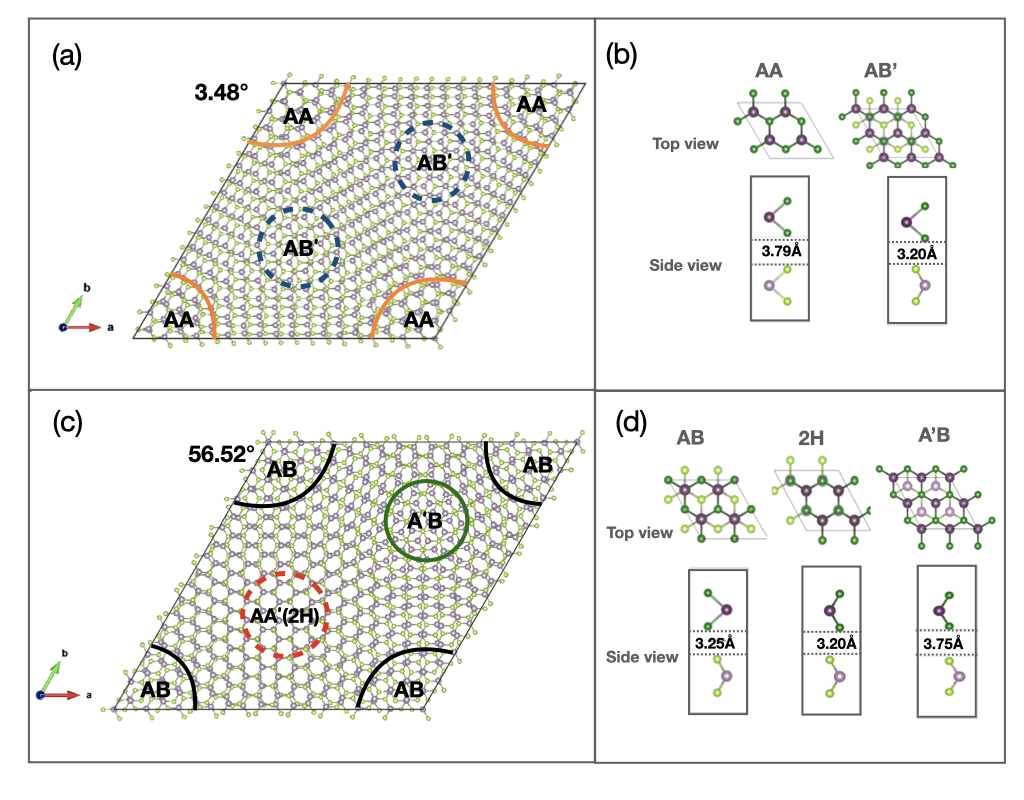}
\caption{Unit cell for twist angle (a) 3.48$^{\circ}$ and (c) 56.52$^{\circ}$. The Mo/Se atoms are shown by violet/green spheres. Regions with high symmetry stackings have been indicated.  The primitive cell stacking AA and AB$'$  }\label{unitboth}
\end{figure}

We have considered the primitive cell of MoSe$_2$ which has AA stacking. While the Mo atoms form a hexagonal lattice, the presence of the Se atoms leads to C$_3$ rotational symmetry instead of C$_6$ rotational symmetry as found in graphene. This makes the rotation about 0$^{\circ}$ and 60$^{\circ}$ inequivalent for TMDCs. For instance, a rotation of 60$^{\circ}$ starting from AA stacking leads to another high symmetry stacking AB, using the notation of identifying primitive cells of Ref.\cite{stacking}. Therefore, we examined the twisted bilayers for twist angles near 0$^{\circ}$ and 60$^{\circ}$ separately. 
For this purpose, we have considered a twist angle of 3.48$^{\circ}$ as well as its conjugate angle 56.52$^{\circ}$ as 56.52$^{\circ}$=(60$^{\circ}$ - 3.48$^{\circ}$). This could also be viewed as twisting by an angle of 3.48$^{\circ}$ clockwise starting from the high symmetry stacking AB. 
Unit cells of both twist angles are shown in Fig. \ref{unitboth}(a) and (b). In each of the cases, the length of each lattice vector of the unit cell is 54.14 $\AA$, with the moir\'{e} cell containing 1626 atoms. For these small twist angles near zero and sixty degrees, one finds large patches of various high symmetry stackings within the moir\'{e} cell. These patches are not found for large twist angles. Even if at a particular point, an atom on atom condition is satisfied, the nearest neighbour atoms are substantially rotated away. This therefore does not allow one to have large regions of any high symmetry stackings for large twist angles. This is the primary distinction between small and large angles of rotation.

Considering the twist angle of 3.48$^{\circ}$ (Fig.~\ref{unitboth}(a)), one finds patches of AA stacking at the corners marked by orange semicircular arcs and patches of AB$'$ stacking in the middle marked by the green circles. The side view and the top view of these primitive cell stackings are shown in Fig. \ref{unitboth}(b).  Depending on the relative position of the atoms of the upper layer with respect to those of the lower layer, each stacking has different interlayer separations. For instance, one finds that in AA stacking, Se atoms of the upper layer are sitting exactly on top of the Se atoms of the lower layer. Hence, the electrons on the Se atoms of the upper layer experience larger Coulomb repulsions from the electrons on the Se atoms of the lower layer. Consequently, the vertical interlayer separation increases in the region of AA stacking and is found to be 3.74 $\AA$. Considering the primitive cell for AA stacking one finds that the equilibrium interlayer separation is similar and is found to be 3.79 $\AA$. In AB$'$ stacking the upper layer is staggered with respect to the lower layer. This is also clear from the top view of the primitive cell AB$'$ stacking (Fig. \ref{unitboth}(b)). The Se atoms of the upper layer sit on top of the centre of the hexagon of the lower layer. Consequently, the electrons on the Se atoms of the upper layer feel a smaller Coulomb repulsion than in the previous case and the layer separation is found to be reduced to 3.19$\AA$ in the region of AB$'$ stacking.  Considering the primitive cell associated with this stacking, one finds the interlayer separation to be 3.20 $\AA$ similar to the patches discussed above. These different stackings can be associated with the domains seen in STEM and AFM experiments.\cite{stem+afm1}.

The twist angles near sixty degrees show large patches of three other high symmetry stackings making it inequivalent to the twist angles near zero degrees. The unit cell for twist angle 56.52$^{\circ}$ is shown in Fig. \ref{unitboth}(c). Here, at the corners, one finds large regions of AB stacking marked with black solid semicircles. In the middle, patches of AA' (marked with red dashed cirle) and patches of A$'$B stackings (marked with green solid circle) are found. The side view and the top view of the primitive cells of these three stackings are shown in Fig. \ref{unitboth}(d). Depending on the orientation of the upper layer with respect to the lower layer, one finds that the that the interlayer separation is substantially different in each case. Considering the primitive cells in each of these three stackings, the equilibrium interlayer separations are found to be 3.25$\AA$, 3.20 $\AA$ and 3.75 $\AA$ for AB, AA'(2H) and A'B stackings respectively. This variation in the interlayer separation has also been reflected in the relaxed moir\'{e} cell of twist angle 56.52$^{\circ}$. The interlayer separation reaches a maximum value of 3.74$\AA$ in the region of A$'$B stacking and reaches a minimum value of 3.20 $\AA$ in the region of AA$'$(2H) stacking.

\begin{figure}[h!]
 \centering
\includegraphics[width=5.5 in]{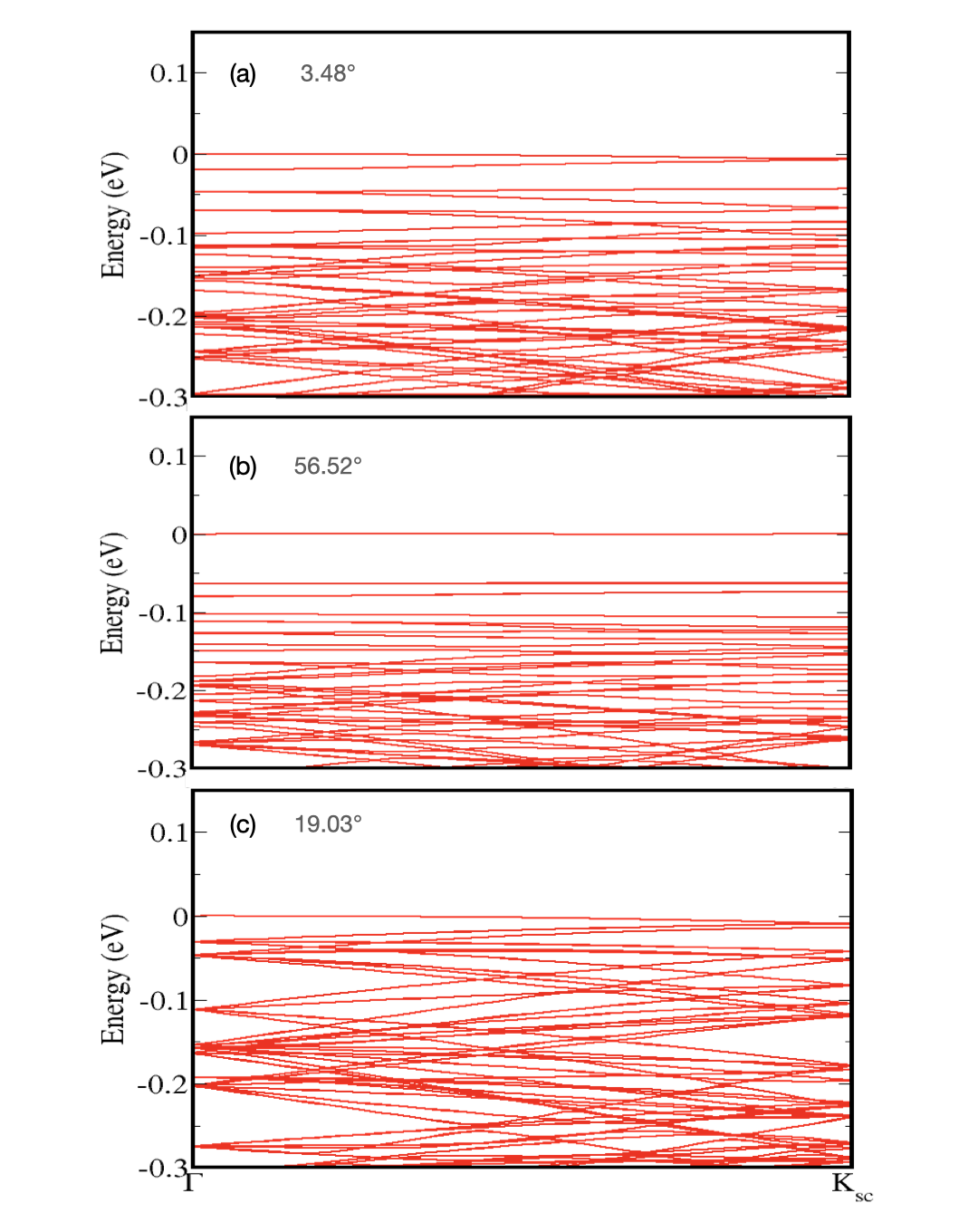}
\caption{Electronic structure (without SOC) for twist angles (a) 3.48$^{\circ}$, (b) 56.52$^{\circ}$ and (c) 19.03$^{\circ}$along $\Gamma$ to K${_sc}$ direction of the supercell.}\label{sc_band}
\end{figure} 

The electronic structure has been calculated for the moir\'{e} cell of twist angle 3.48$^{\circ}$ as well as 56.52$^{\circ}$. 
However, as a consequence of the large moir\'{e} cells, the electronic structure gets folded into very small Brillouin zone resulting in flat dispersionless bands irrespective of the angle of twist.  The folded electronic structure plotted in the small Brillouin zone associated with the moir\'{e} along the $\Gamma$ to K$_{sc}$ direction, for twist angles 3.48$^{\circ}$, 56.52$^{\circ}$ as well as 19.03$^{\circ}$ have been shown in Fig. \ref{sc_band}. In each of the cases, we find  almost dispersionless bands in the entire energy window, which is contrary to experiments. We have discussed this aspect in Ref.~ \cite{twist_mose2_pk,twist_mose2_sp,twist_wse2_sp}, explaining why the interesting phenomena are seen in the small
angle regime. 

\begin{figure}[h!]
 \centering
\includegraphics[width=5.8 in]{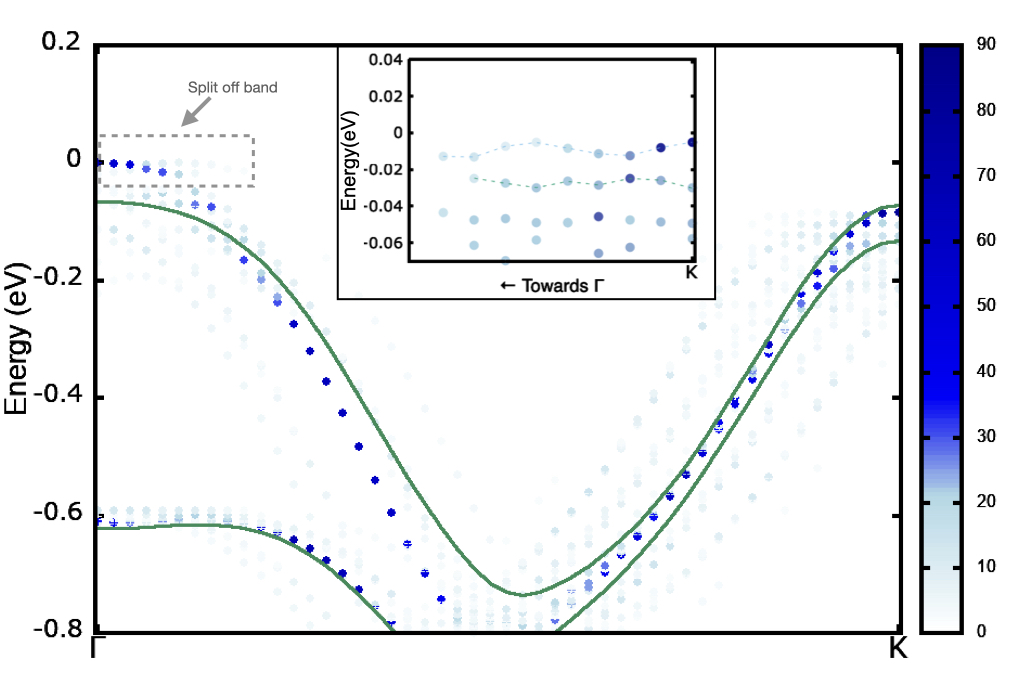}
\caption{Electronic structure (without SOC) for twist angle 3.48$^{\circ}$ projected along the primitive cell $\Gamma$ to K direction, with the colour representing the weight at that k-point. The green line shows the primitive cell band
structure for AA stacking at an interlayer separation of 3.49 $\AA$. An expanded view of the highest
occupied bands near K including SOC is shown as an inset. }\label{3p48band}
\end{figure}

As the moir\'{e} cell electronic structure can be viewed as a weak perturbation to the electronic structure of the untwisted limit,
each eigenfunction of the moir\'{e} cell  has been projected onto the eigenfunction of the primitive cell along the $\Gamma$ to K direction, as described in the methodology section. The projected band structure along the $\Gamma$ to K of the primitive cell is shown in Fig. \ref{3p48band} for twist angle 3.48$^{\circ}$. The colour bar on the right side represents the percentage of the weight of the energy eigenstate at that primitive cell kpoint.

In contrast to the band dispersions expected for a moir\'{e} supercell, where one expects almost dispersionless bands, here, we found a set of dispersing bands. These dispersing bands are similar to what is seen in the untwisted limit. The untwisted limit band structure is superposed in Fig. \ref{3p48band} for comparision. One finds that the projected eigenfunctions which have large weights, follow the band dispersions of the untwisted limit. However in addition to that, a weakly dispersing band emerges at the top of the valence band at $\Gamma$. This split-off band is well separated from the valence bands below and has a bandwidth of 19 meV. 
The highest occupied band at K is found to be energetically much lower than that at  $\Gamma$ with the separation being 84 meV. Including spin orbit interactions reduces this $\Gamma$-K separation to 5 meV with the valence band maximum remaining at 
$\Gamma$. An expanded view of the near K region of the band structure close to the fermi energy,
calculated including spin-orbit interactions is shown in the inset of Fig. \ref{3p48band}. This
indicates that there are flat bands at K.

\begin{figure}[h!]
\centering
\includegraphics[width=5.8 in]{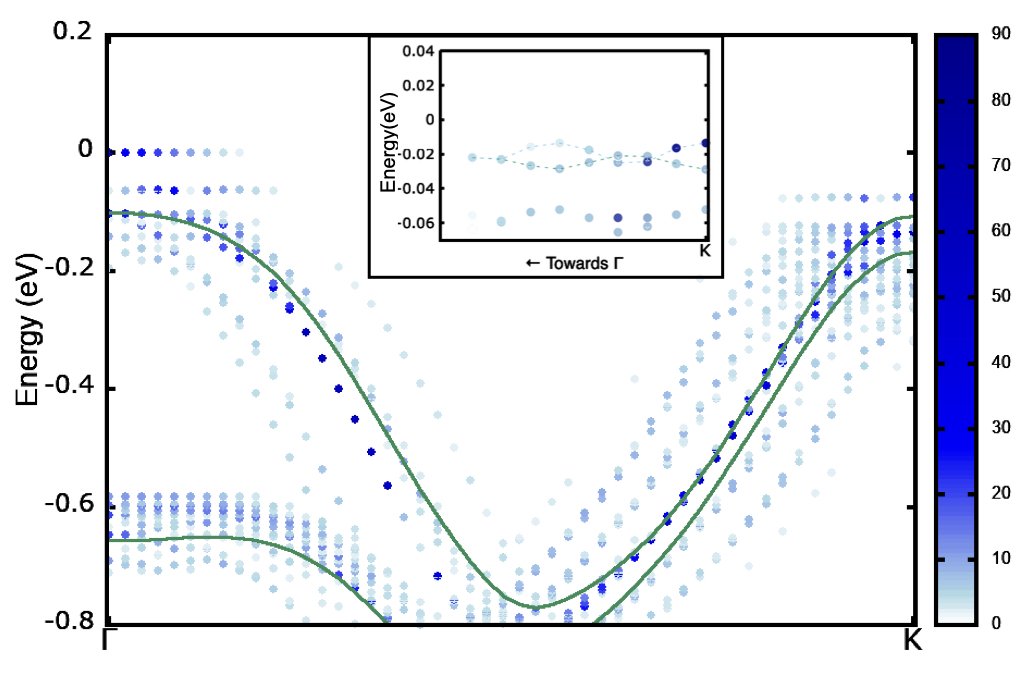}
\caption{Electronic structure for twist angles 56.52$^{\circ}$ projected along the primitive cell $\Gamma$ to K direction, with the colour representing the weight at that k-point. The green line shows the primitive cell band structure for AA stacking at an interlayer separation of 3.49 $\AA$. An expanded view of the highest
occupied bands near K including SOC is shown as an inset.}\label{56p52band}
\end{figure}

In a similar manner, the projected band structure for twist angle 56.52$^{\circ}$ along the $\Gamma$ to K direction of the primitive cell has been plotted in Fig. \ref{56p52band} with the untwisted limit band structure superposed. Here also, we found a set of dispersing bands associated with the valence bands, like the untwisted limit. In addition to those, a pair of split off bands are found above the dispersing valence bands. These split-off bands have a vanishingly small bandwidth. 
Here also, the highest occupied  valence band at K is energetically much lower than that at  $\Gamma$ with the $\Gamma$-K separation being 74 meV. Inclusion of spin orbit 
interactions reduces this $\Gamma$-K separation to 13 meV with $\Gamma$ being the valence band maxima. An expanded view of the
near K region of the band dispersions close to the fermi energy, calculated including spin-orbit interactions has been shown in 
Fig. \ref{56p52band}. One finds flat bands emerging 
at K also.

However, these split-off bands are not found for large twist angles\cite{twist_mose2_sp} and the band structure is similar to the untwisted limit.  It immediately raises the question what is the origin of the perturbation in the electronic structure of small twist angles near 0$^{\circ}$ and 60$^{\circ}$.

In these materials, the layers are coupled to each other by weak van der Waals forces. In a previous work\cite{shishir}, we have examined the origin of the changes in the  electronic structure with layers. 
This has been carried out by mapping the ab-initio electronic structure of a bilayer primitive cell onto a tight binding model. Within this model, the interlayer hopping interactions  have been switched off. This gives us  the monolayer's bandstructure. This implies that the changes in the electronic structure is due to the interlayer hopping interactions only.

\begin{figure}[h!]
 \centering
 \includegraphics[width=6 in]{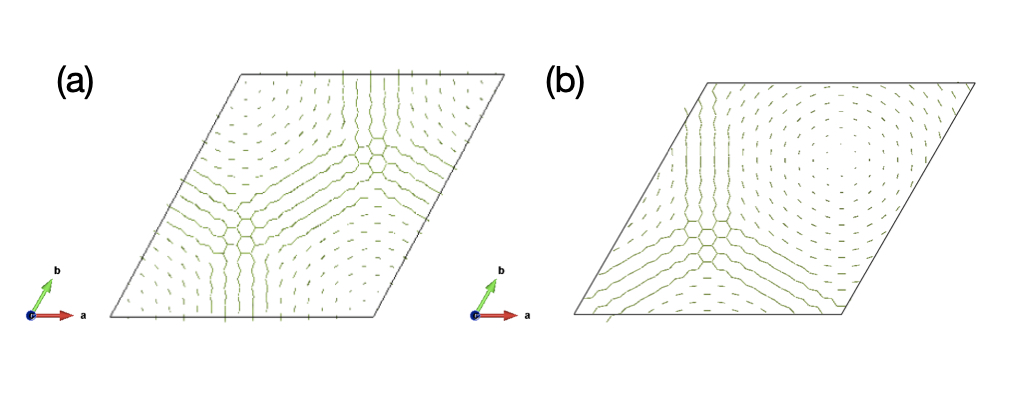}
 \caption{ Spatial profile of interlayer Se-Se distances less than 3.8$\AA$ twist angle (a) 3.48$^{\circ}$ and (b) 56.52$^{\circ}$}\label{distprofile}
\end{figure}

Therefore, in order to understand the origin of the perturbation in case of the small twist angles near zero and sixty degrees, one needs to examine the interlayer hopping interactions in these moire cells. 
As these scale with the
interlayer Se-Se distance, we plotted the spatial profile of the interlayer Se-Se bond lengths which are less than 3.8$\AA$ in each of the cases. This is shown in \ref{distprofile} (a) and (b) for twist angles 3.48$^{\circ}$ and 56.52$^{\circ}$ respectively. The regions where these shortest bond lengths are concentrated is around the region of AB$'$ stacking for twist angle 3.48$^{\circ}$ as shown in Fig. \ref{distprofile} (a) and around the region of 2H stackings for twist angle 56.52$^{\circ}$ as shown in Fig. \ref{distprofile} (b). In these regions, the Se atoms of one layer have more number of Se neighbours from the other layer with these bond lengths.
Hence these regions give rise to a strong perturbation to the electronic structure of the untwisted limit in each case. 

Examining the charge density associated with the highest occupied band, the untwisted limit has the charge density almost uniformly  distributed on both layers. However, at these small 
twist angles the 
spatial distribution of the interlayer hopping interaction strengths modifies the charge density localization. The unit cell 
as well as the localization of the charge density for a twist angle 
of 3.48$^{\circ}$ is shown in 
Fig. \ref{distprofile}(a). It follows the spatial profile of the enhanced interlayer hopping interactions and is localized in the same region. A
similar localization trend is also seen for the twist angle of 56.52$^{\circ}$. 

\begin{figure}[h!]
 \centering
 \includegraphics[width=6 in]{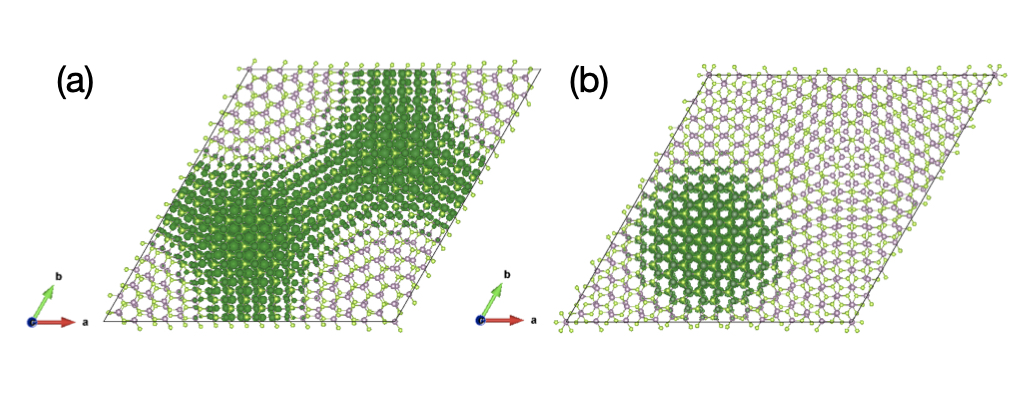}
 \caption{Charge density associated with the highest occupied split off band at $\Gamma$ for twist angle (a) 3.48 $^{\circ}$ and (b) 56.52 $^{\circ}$}\label{chg}
\end{figure}

\begin{figure}[h!]
 \centering
 \includegraphics[width=4 in]{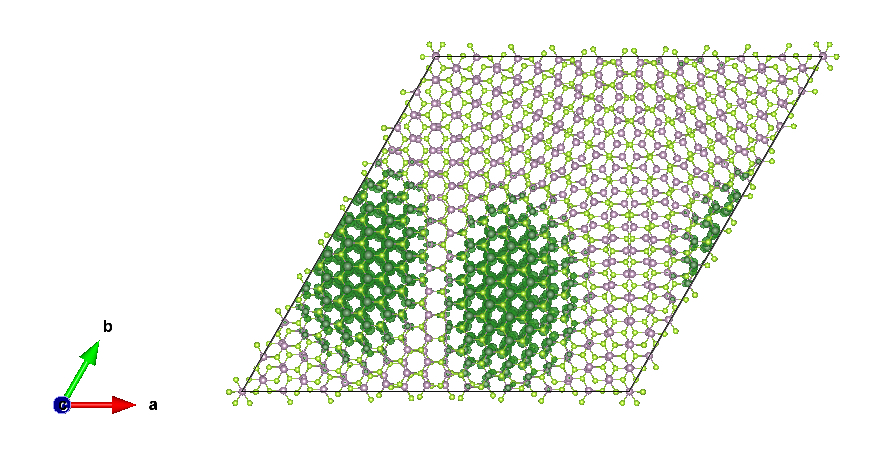}
 \caption{Charge density associated with the second highest split off band at $\Gamma$ for twist angle 56.52 $^{\circ}$}\label{chgden}
\end{figure}

The hopping interaction between
the orbitals on the two layers leads to a set of bonding and antibonding states. One can view 
the charge density plotted in Figs. \ref{chg}(a) and (b) as being associated with the antibonding state emerging 
from the region with enhanced interaction between the two layers.   
The stronger interactions also pushes the antibonding state to 
higher energies compared to the state in the region with weaker interactions. One can 
therefore associate a
quantum well of spatial width L in the region associated with each of the two lobes of the charge density in Fig. 5(a), with a small barrier in between.
A similar potential well
of width L can be associated with the 
region where 
the charge density is localized in Fig. \ref{chg}(b).
As the interlayer separations in the regions of the well should be similar, one expects potential wells
with similar depths, say V$_0$, in both cases.
Associating the flat bands that we find with the bound states of the 
potential well problem,
we can try to qualitatively understand the existence of flat bands at 3.48$^{\circ}$ and
56.52$^{\circ}$. Let the well
of width L have two bound states for the potential V$_0$. Approximating the barrier width to zero in the scenario where the  two potential wells merge, their width is now 2L. As the energy of each state
varies as $\frac{1}{(2L)^2}$, the lowest energy bound state is now shallower, and
so we have only one flat band here. This
qualitative analysis helps us understand the origin of one flat band for 3.48$^{\circ}$, and the existence of two flat bands for 56.52$^{\circ}$,
with the energetically deeper flat band being 
associated with the 
excited state of the
potential well problem.
This is evident from the 
charge density shown in Fig. \ref{chgden}. 

\begin{figure}[h!]
 \centering
 \includegraphics[width=4.4 in]{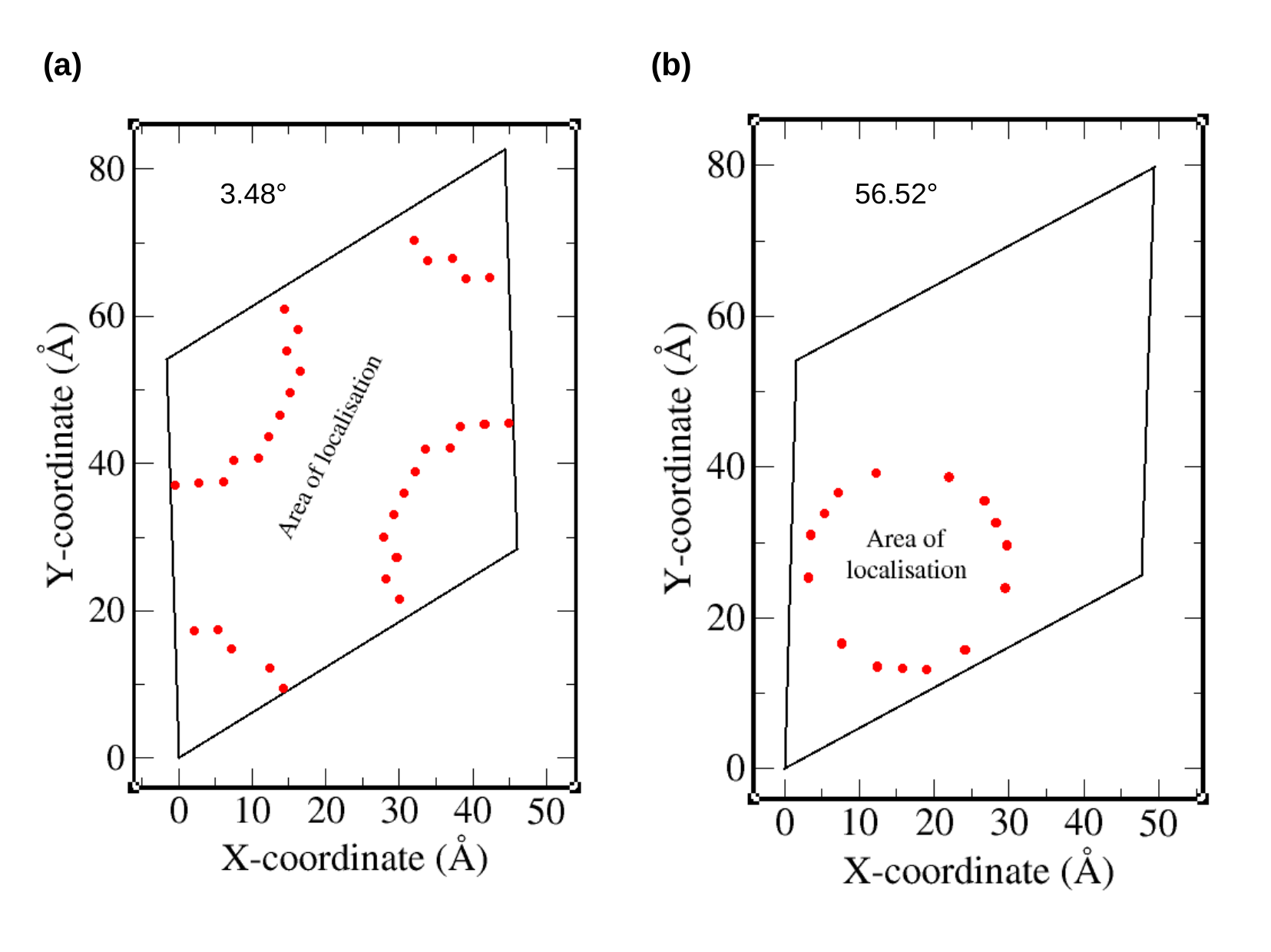}
 \caption{The area of the charge density localisation of the split off band up to where the integrated charge density over a sphere of radius 1.262 $\AA$ (half the Mo-Se bondlength) around each Mo atom falls to 10$\%$ of its maximum value for twist angle (a)3.48$^{\circ}$ and (b)56.52$^{\circ}$} \label{chgdenprof}
\end{figure}

Examining the dispersional width of the band, one finds it to be 19~meV for a  twist angle of 3.48$^{\circ}$, while it is vanishingly small for 56.52$^{\circ}$. In order to understand this, we have plotted the area of localisation of the charge density of this band in each case.  This has been determined by calculating the integrated charge density around each Mo atom by considering a sphere of a radius of 1.262$\AA$ (half of Mo-Se bond length). We have marked the area of the charge density localisation up to where the integrated charge density over a sphere around each Mo atom falls to 10 \% of its maximum value for both the twist angles 3.48$^{\circ}$ and 56.52$^{\circ}$ in Fig. \ref{chgdenprof} (a) and (b) respectively.

It is evident from Fig.\ref{chgdenprof} that the area of charge density localisation of the split off band is significantly smaller for twist angle 56.52$^{\circ}$ in comparison to 3.48$^{\circ}$. This implies that the smaller dispersional width of a band or constant energy in k-space could be associated with more localized wave function in real space. 

In conclusion, we have examined the electronic structure of small rotations about AA and AB stacking. This leads to 
a low energy electronic structure involving dispersive bands, quite similar to the untwisted limit. In addition we see almost flat bands just above these dispersing bands. We find that the number of flat bands forming, and their localization can be discussed in terms of the spatial extent of the perturbing potential. \\ \\ \\ \\ \\

The authors acknowledge the support from SERB through Project No. IPA/2020/000021. The support and the resources provided by PARAM Shivay Facility under the National Supercomputing Mission, Government of India at the Indian Institute of Technology, Varanasi are gratefully acknowledged. The authors also acknowledge the us of the Technology Research Centre facilities at S.N. Bose centre. MD acknowledges a fellowship from the DST-INSPIRE (IF200577), Government of India  .



\clearpage

\begin{thebibliography}{40}

\bibitem{tw_graphene1}
Y. Cao, V. Fatemi, S. Fang, K. Watanabe, T. Taniguchi, E. Kaxiras, and P. Jarillo-Herrero, Nature (London) \textbf{556}, 43 (2018).

\bibitem{tw_graphene2}
Y. Cao, V. Fatemi, A. Demir, S. Fang, Spencer L. Tomarken, J. Y. Luo, J. D. Sanchez-Yamagishi, K. Watanabe, T. Taniguchi, E. Kaxiras, R. C. Ashoori, P. Jarillo-Herrero, Nature (London) \textbf{556}, 80 (2018).

\bibitem{tw_graphene3}
Y. Jiang, X. Lai, K. Watanabe, T. Taniguchi, K. Haule, J. Mao, and E. Y. Andrei, Nature (London) \textbf{573}, 91 (2019).

\bibitem{tw_graphene4}
S. Wu, Z. Zhang, K. Watanabe, T. Taniguchi and E. Y. Andrei, Nat. Mat. \textbf{20}, 488 (2021).

\bibitem{tw-WSe2_1}
L. Wang, E.-M. Shih, A. Ghiotto, L. Xian, D. A. Rhodes, C. Tan, M. Claassen, D. M. Kennes, Y. Bai, B. Kim, K. Watanabe, T. Taniguchi, X. Zhu, J. Hone, A. Rubio, A. Pasupathy, and C. R. Dean, Nat. Materials \textbf{19}, 861 (2020).

\bibitem{tw-WSe2_2}
A. Ghiotto, E. -M. Shih, G. S. S. G. Pereira, D. A. Rhodes, B. Kim, J. Zang, A. J. Millis, K. Watanabe, T. Taniguchi, J. C. Hone, L. Wang, C. R. Dean and A. N. Pasupathy, Nature \textbf{597}, 345 (2021). 

\bibitem{tw_graphene5}
L. A. Gonzalez-Arraga, J. L. Lado, F. Guinea, and P. San-Jose, Phys. Rev. Lett. \textbf{119}, 107201 (2017).

\bibitem{tw-MoSe2_1}
P.-C. Yeh, W. Jin, N. Zaki, J. Kunstmann, D. Chenet, G. Arefe, J. T. Sadowski, J. I. Dadap, P. Sutter, J. Hone, and J. R. M. Osgood, Nano Lett. \textbf{16}, 953 (2016).

\bibitem{alloy+doped}
V. Popescu, and A. Zunger, Phys. Rev. B \textbf{85}, 085201 (2012).

\bibitem{alloy_unfold}
Y. Chen, J. Xi, D. O. Dumcenco, Z. Liu, K. Suenaga, D. Wang, Z. Shuai, Y.-S. Huang, and L. Xie, ACS Nano, 7, 4610 (2013).

\bibitem{doped_unfold}
M. Reticcioli, G. Profeta, C. Franchini, and A. Continenza, Phys. Rev. B \textbf{95}, 214510 (2017).

\bibitem{twist_mose2_pk}
P. Kumari, J. Chatterjee and P. Mahadevan, Phys. Rev. B \textbf{101}, 045432 (2020).

\bibitem{twist_mose2_sp}
S. Patra, P. Kumari and P. Mahadevan, Phys. Rev. B \textbf{102}, 205415 (2020).

\bibitem{twist_wse2_sp}
S. Patra, P. Boyal and P. Mahadevan, Phys. Rev. B \textbf{107}, L041104 (2023).

\bibitem{tw_theo1}
M. H. Naik and M. Jain, Phys. Rev. Lett. \textbf{121}, 266401 (2018).

\bibitem{tw_theo2}
M. H. Naik, S. Kundu, I. Maity, and M. Jain, Phys. Rev. B \textbf{102}, 075413 (2020).
\bibitem{continuum01}
Proceedings of the National Academy of Sciences, R. Bistritzer and A. H. MacDonald,  \textbf{108}, 12233 (2011).
\bibitem{continuum02}
S. Javvaji, J.-H. Sun, and J. Jung. Phys. Rev. B \textbf{101}, 125411 (2020).
\bibitem{continuum03}
T. Devakul, V. Crepel, Y. Zhang and L. Fu, Nature Communications, \textbf{12}, 6730 (2021).
\bibitem{vasp-1}
G. Kresse and J. Hafner, Phys. Rev. B \textbf{47}, 558 (1993).

\bibitem{vasp-2}
G. Kresse and J. Hafner, Phys. Rev. B \textbf{49}, 14251 (1994).

\bibitem{paw}
P. E. Blöchl, Phys. Rev. B \textbf{50} , 17953 (1994).

\bibitem{pbe-1}
J. P. Perdew, K. Burke, and M. Ernzerhof, Phys. Rev. Lett.\textbf{77}, 3865 (1996).

\bibitem{pbe-2}
J. P. Perdew, B. Kieron, and E. Matthias, Phys. Rev. Lett. \textbf{78}, 1396 (1997).

\bibitem{dft-d2}
S. Grimme, J. Comput. Chem. \textbf{27}, 1787 (2006).

\bibitem{unfold}
V. Popescu and A. Zunger, Phys. Rev. B \textbf{85}, 085201 (2012).

\bibitem{wavetrans}
R. M. Feenstra, N. Srivastava, Q. Gao, M. Widom, B. Diaconescu, T. Ohta, G. L. Kellogg, J. T. Robinson, and I. V. Vlassiouk, Phys. Rev. B \textbf{87}, 041406(R) (2013).

\bibitem{stacking}
J. He, K. Hummer, and C. Franchini, Phys. Rev. B \textbf{89}, 075409 (2014). 

\bibitem{stem+afm1}
A. Weston, et al. Nature Nanotech. \textbf{15}, 592 (2020).

\bibitem{shishir}
S. K. Pandey, R. Das, and P. Mahadevan, ACS Omega \textbf{5}, 15169 (2020)


\end{thebibliography}
\end{document}